\documentstyle[12pt,epsf]{article}
\title{\bf Abundances in Very Metal-Poor Stars}
\author{S. G.~Ryan \\
\vspace{1cm}\\
\normalsize Physics Dept, The Open University, MK7 6AA, United Kingdom\\}
\date{\mbox{}}
\begin{document}
\maketitle
\pagestyle{empty}
%
% WE REDEFINE THE plain LaTeX PAGESTYLE !!! 
% THIS PAGESTYLE WILL BE USED FOR THE FIRST PAGE ONLY !
%
\def\bull{\vrule height .9ex width .8ex depth -.1ex}
\makeatletter
\def\ps@plain{\let\@mkboth\gobbletwo
\def\@oddhead{}\def\@oddfoot{\hfil\tiny\bull\quad
``The Galactic Halo: from Globular Clusters to Field Stars'';
35$^{\mbox{\rm th}}$ Li\`ege\ Int.\ Astroph.\ Coll., 1999\quad\bull}%
\def\@evenhead{}\let\@evenfoot\@oddfoot}
\makeatother
%
% AND DEFINE OUR MACROS FOR THE REFERENCE LIST
% I.E \beginrefer \refer and \endrefer
%
\def\beginrefer{\section*{References}%
\begin{quotation}\mbox{}\par}
\def\refer#1\par{{\setlength{\parindent}{-\leftmargin}\indent#1\par}}
\def\endrefer{\end{quotation}}
%
% BEGIN THE ABSTRACT CHAPTER WITH \noindent\small, ENCLOSE IT IN A GROUP
% AND BOLDFACE THE TITLE.
%
{\noindent\small{\bf Abstract:} 
Metal-poor stars provide information on the characteristics and chemical 
evolution of the halo population of the
Galaxy, the first epoch of
 
star formation and Galaxy formation (not just locally but with relevance to
high-redshift objects), and big bang nucleosynthesis.
This review looks at recent developments in this subject. 

}
%
% NOW COMES THE MAIN BODY OF THE ARTICLE
%
\section{Introduction}

Halo stars can be viewed in several contexts. They constitute the oldest and 
most extended stellar population {\it known} in the Galaxy.
As probes of Galactic chemical evolution (GCE), they 
are
the oldest objects and have the lowest metallicities, and hence 
provide the first data in the evolutionary sequence. 
In a third context, the early
evolution of the universe, Figure 1 shows the metallicity distributions of
36 halo globular clusters (Laird et al. 1988a), 373 halo field stars 
(Laird, Carney \& Latham 1988b; Ryan \& Norris 1991)
and 34 damped Lyman-$\alpha$ systems (DLAs;  Pettini et al. 1997).
Not only are the field and cluster metallicity
distributions comparable, they are {\it lower} in metallicity than the DLAs 
having redshifts $z \sim$~1--3. That is, very metal poor stars 
are amongst the lowest metallicity objects in the known universe.

%% \begin{figure}
%% \vspace{5 cm}
%% \caption{Metallicity distributions of Galactic halo globular clusters, 
%% Galactic halo field stars, and Damped Lyman-$\alpha$ quasar absorption line 
%% systems.}
%% \end{figure}

\begin{figure}[!htb]
\begin{center}
\leavevmode
\epsfysize=130mm
\epsfbox[193 168 403 569]{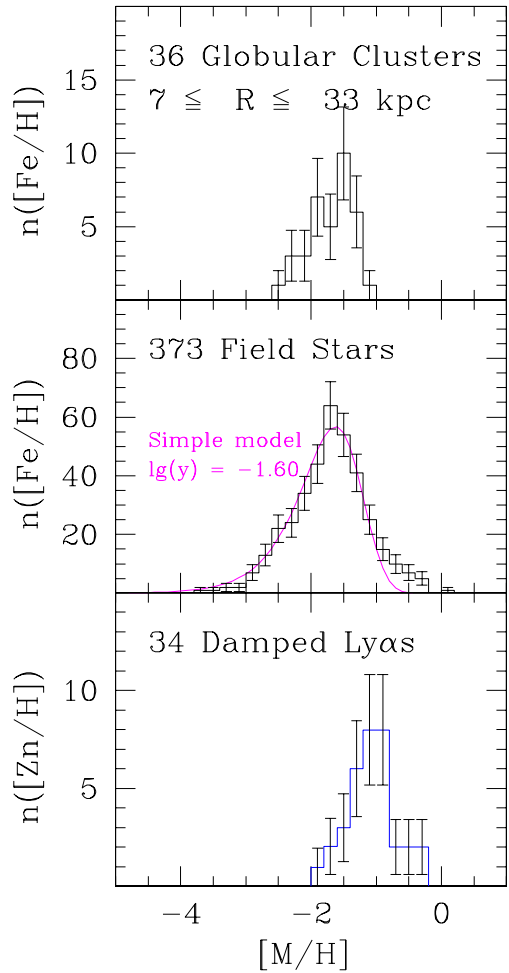}
% \epsfbox{liege_fig1.ps}
\end{center}
% % psfile=#1 vsize=#2 angle=#3 hscale=#4 vscale=#5 hoffset=#6 voffset=#7
% \plotfiddle{sample_fig1.eps}{75truemm}{0}{45}{45}{-150}{-80}
\caption{Metallicity distributions of Galactic halo globular clusters, 
Galactic halo field stars, and Damped Lyman-$\alpha$ quasar absorption line 
systems.}
\end{figure}

The surprise some people express in discovering that DLAs are
 generally 
more metal rich than the Galactic halo emphasises that our
knowledge of the DLAs has yet to mature. There is ongoing
debate about what they really are, 
possibilities including:
\newline $\bullet$ spiral disks/protodisks/thick disks (e.g. Wolfe et al. 1986;
Lu et al. 1996)
\newline $\bullet$ dwarf galaxies (Pettini, Boksenberg \& Hunstead 1990; Pettini
et al. 1999a)
\newline $\bullet$ ejecta from dwarf galaxies (Nulsen, Barcons \& Fabian 1998).
\newline
In examining Galactic stars, we have the advantage of studying objects with
reasonably well understood histories and physical states, whose
spectra are not blended, and whose abundances, which are 
measurable for many elements,
are unaffected by depletion onto dust grains.

The value of halo stars for studying early epochs 
of the universe may be further illustrated by considering 
additional objects in
metallicity-redshift space.  Figure 2 shows a number of Galactic and 
high-redshift objects, along with three GCE models assuming 
outflow, no outflow (closed box) and inflow for Galaxy formation assumed to 
begin at redshift $z = 5$ (Edmunds \& Phillipps 1997). (A Hubble constant 
$H_0 = 50$~km~s$^{-1}$~Mpc$^{-1}$ and flat cosmology [deceleration parameter
$q_0$ = 0.5] were used to establish the age-redshift relation.)
The disk star sequence and bulge of the Galaxy are 
based on the observations of Edvardsson et al. (1993)
and Sadler, Rich \& Terndrup (1996). 
The high-redshift objects are the DLAs from
Pettini et al. (1997) binned in redshift, Molaro et al.
(1996) and Lu et al. (1996), a region corresponding to the Lyman-$\alpha$ 
forest (Hellsten et al. 1997), and one Lyman break galaxy (LBG) from 
Pettini et al. (1999b). 
The redshift distribution in the Hubble Deep Field (Bouwens, Broadhurst \& Silk 
1998) is shown as an inset against the vertical axis.
In adjoining panels are the
extinction-corrected star formation rate 
(Steidel et al. 1999) and the quasar space density 
(Warren, Hewitt \& Osmer 1994) whose steep fall at
redshift $z$~$>$~3.5 indicates that these objects were still 
forming prior to this epoch, presumably along with galaxies.
Several points can be noted. 
\newline $\bullet$ The occurrence of systems covering a wide metallicity
range at the same high redshift --- 
the Lyman-$\alpha$ forest at $-3$~$<$~[M/H]~$<$~$-2$,
DLAs at $-2$~$<$~[M/H]~$<$~$-1$,
and a Lyman break galaxy at [M/H]~$>$~$-1$ ---
suggests that we are probing diverse objects, not necessarily an 
evolutionary sequence, in the high redshift universe. 
\newline $\bullet$ The overlapping of high redshift and 
Galactic objects in
this epoch-metallicity plane (redshift translates to epoch),
e.g. the LBG and the Galactic bulge,
and the Lyman-$\alpha$ forest and metal poor stars, 
emphasises that these objects provide complementary views
on the formation and evolution of galaxies. No one class should be 
considered in
isolation from the others.
\newline $\bullet$ Galactic stars with [Fe/H]~$<$~$-3$, corresponding to 
$z$~$^>_\sim$~4--5,
uniquely probe the 
earliest star formation events. A high level of 
detail is achievable because many elements can be
measured in well understood objects. Furthermore, 
the elements in these objects owe their existence to very few 
previous generations of stars, possibly only one (Ryan, Norris \& Bessell 1991).
They mark the beginning of GCE,
and as such will be the main topic of this review.

%% \begin{figure}
%% \vspace{5 cm}
%% \caption{Metallicity distributions of Galactic halo globular clusters, 
%% Galactic halo field stars, and Damped Lyman-$\alpha$ quasar absorption line 
%% systems.}
%% \end{figure}

\begin{figure}[!htb]
\begin{center}
\leavevmode
\epsfxsize=167mm
\epsfbox[46 168 564 430]{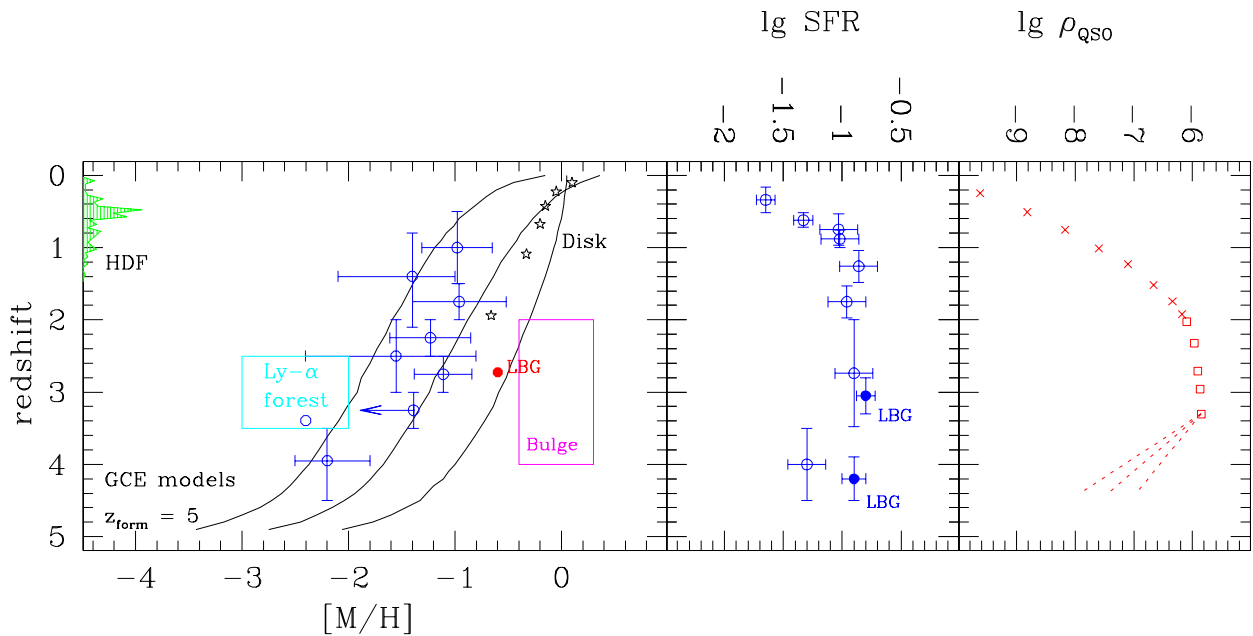}
% \epsfbox{liege_fig2.ps}
\end{center}
% % psfile=#1 vsize=#2 angle=#3 hscale=#4 vscale=#5 hoffset=#6 voffset=#7
% \plotfiddle{sample_fig1.eps}{75truemm}{0}{45}{45}{-150}{-80}
\caption{(a): Galactic and high-redshift objects in the 
redshift--metallicity plane, the former located at their
redshift of formation (assuming $H_0 = 50$~km~s$^{-1}$~Mpc$^{-1}$).
{\it solid curves} = GCE models;
{\it open circles} = DLAs; 
{\it stars} = disk star sequence; 
{\it solid circle} = LBG.
(b) Star formation history of extragalactic objects.
(c) Quasar density distribution.
(See text for discussion.)}
\end{figure}

\section{How Many Supernovae Make a Population II Star?}

Disk stars formed from gas
enriched by many previous generations of stars. 
Population II stars formed earlier and have lower metallicities, consistent 
with fewer supernovae being
involved.  A simple closed box model of GCE
(e.g. Searle \& Sargent 1972; Pagel \& Patchett 1975) establishes a  
framework based on a single free parameter --- the fraction of enriched matter 
returned to the
interstellar medium by a stellar generation, known as the yield. 
No check is kept on timescales or stellar generations; with
instantaneous processing and complete mixing, the model follows essentially the
average evolution of a galaxy. Despite these gross simplifications, the
model compares very favourably with the metallicity distribution
of the Galactic halo; Figure 1 compares halo field star data 
with a simple model with yield $y = 10^{-1.6}$ (Ryan \& Norris 1991).
However, models of this type are incapable of indicating how many supernovae
are required to enrich a stellar Population to a given metallicity, and 
alternatives treating the formation and evolution of stars
more realistically had to be developed. 

One model which considered 
individual generations of stars was Searle's (1977) stochastic model, which
postulated that separate star forming fragments underwent star formation
according to Poisson statistics, the mean number of enrichment events
increasing with time
and the enrichment from each being constant. This model again has
only one free parameter, the mean number of enrichments 
prior to the termination of star formation, $\mu$, but it is capable of
quantifying the number of supernovae involved.
Its metallicity distribution 
is broadly similar to that of the simple model.  Applied to 
halo field stars, Searle's stochastic model was not obviously
better than the simple model, but did give a 
marginally better fit to 
the globular cluster distribution, implying a mean of ten
enrichments per fragment (Ryan \& Norris 1991). With the mean metallicity of 
the halo globular 
clusters being [Fe/H]~$\simeq$~$-1.5$, this suggested that a single event
could enrich a fragment to [Fe/H]~$\simeq$~$-2.5$.

Truran (1981) argued that the atmospheres of second generation stars would 
contain r- but no s-process isotopes, due to the latter's secondary nature, 
whereas subsequent generations would contain both types. 
Although it was not clear at which metallicity this 
would be achieved, it forced people to think about the 
first few generations of stars.\footnote{
It is uncertain that we will actually see a clear division between
r- and s-process elements in second versus third generation stars, because
most neutron-capture elements 
have contributions from both the s- and r-process.}
Large
star-to-star variations in the neutron-capture abundances of stars with
[Fe/H]~$<$~$-2.5$
(e.g. Gilroy et al. 1988) suggested that prior nucleosynthesis involved small
numbers of supernovae (SN).
Related observations of Sr
led Ryan et al. (1991) to 
consider stochastic enrichment by
only a few stellar generations, and to 
calculate the
metallicity produced by single supernova as 
[Fe/H]~=~$-3.8$,
based on a typical SN II progenitor mass of 25~M$_\odot$ and an assumed
primordial cloud mass of $10^6~$M$_\odot$.\footnote{The SN mass was a compromise 
between higher mass stars being rarer and lower mass stars having
lower yields.  The cloud mass was based on large globular clusters,
giant molecular clouds, and the collapse of metal poor gas.}
This coincided 
with the observed onset of
huge abundance variations of strontium (by a factor of 100 or more), and
was consistent with the lowest stellar metallicities then 
observed.\footnote{The vanishing of Sr in some stars at this metallicity 
suggested to me, at the time, that Truran's mechanism was possibly being
observed, and that genuine second generation stars were being identified. 
However, the more recent availability of data on Ba has altered my views on
this; see Footnote 1 and later sections of this paper.}

Star-to-star differences in neutron-capture element abundances at the lowest 
metallicities also required
that chemical inhomogeneities existed around the time these stars were forming. 
Audouze \& Silk (1995) showed further that
mixing timescales in the halo were sufficiently long 
that inhomogeneities of this type would not be erased on the timescale over
which stars would form, supporting the proposition that 
the progeny of the first supernovae
would be found around this metallicity. Mounting examples of 
neutron-capture element variations from star to star (e.g. Norris, 
Peterson \& Beers 1993; McWilliam et al.
1995; Ryan 1996; Ryan et al. 1996) strengthened the view that the most metal-poor stars 
exhibit the ejecta of very small numbers of
supernova.

The need to integrate small number statistics of the first supernova 
with GCE
models led Ryan, Norris \& Beers (1996) to examine the enrichment sphere of 
a single SN in a primordial cloud. Adopting the supernova remnant (SNR)
model of Cioffi, 
McKee \& Bertschinger (1988), they calculated the cloud mass with
which the SN ejecta would mix as
$$m_{\rm ISM} = 3.4\times10^4 E_{51}^{0.95} n_0^{-0.10}\zeta_m^{-0.15} 
(\beta C_{06})^{-1.29}~~~~{\rm M}_\odot,$$
where $E$, $n$, $\zeta$, and $\beta C_{06}$ refer to the explosion energy,
cloud density, cloud metallicity, and shock speed in appropriate units.
The main features of this result were:
\newline $\bullet$ the mass of gas enriched is almost
independent of the cloud characteristics ($n, \zeta$)
and depends strongly
(almost linearly) on the energy of the SN;
\newline $\bullet$ the typical enriched mass of the cloud would be 
$7\times10^4~$M$_\odot$; 
\newline $\bullet$ the typical metallicity following this first enrichment
would be [Fe/H]~=~$-2.7$, matching (perhaps coincidentally) the changes in the 
behaviour of iron-peak and neutron-capture elements and 
the lowest metallicity globular clusters.

Many other GCE models have been developed that
incorporate the initial mass function, the 
mass-dependence of stellar lifetimes, the mass-
and metallicity-dependence of supernova and stellar-wind yields, 
and more. In the light of
observational and theoretical reasons for expecting
the first star forming regions to be poorly mixed, 
new GCE models have been forthcoming
that include SNR physics and
inhomogeneous mixing (e.g. Shigeyama \& Tsujimoto 1998; Tsujimoto \& Shigeyama
1998; Ishimaru \& Wanajo 1999a; Argast \& Samland 1999),
against which the 
abundances of very metal poor stars can be compared. 
It is then possible to invert the problem and use the
observed abundances to constrain the model inputs. As stars at these low
abundances are believed
to be second generation stars, of particular interest will
be the shape of the IMF of their progenitors (Population III stars!),
the mass limits for the production of SN of Population III stars,
and the yields of individual Population III objects.

\section{Abundances: Can You Believe What You Read?}

Weak lines have the greatest
sensitivity to abundance and the least sensitivity to uncertain parameters 
of the stellar atmosphere. However, the lines that are weak in very
metal poor stars are strong in the
sun, so completely different lines are measured in the two cases,
the former often also being of lower excitation potential and
possibly of a different ionisation state.  Photometric temperature 
calibrations and stellar atmosphere models also depend 
on metallicity.
These factors give rise to potential systematic differences between 
analyses
conducted for metal rich compared to metal poor stars. The overall rarity of
spectral lines in metal poor stars also limits consistency checks 
between several lines of one element.
There can also be substantial
differences in the approaches of investigators, who may adopt
different reference solar abundances, model atmosphere grids, and atmospheric 
parameters. 
Differences of 10\% in the equivalent widths measured in two 
studies of one star are not uncommon.
Differences of 0.2~dex in the
absolute abundances can accumulate through these effects.

Fortunately, 
for many species relative abundances (ratios) 
[X/Fe] are less susceptible to 
errors than [X/H]; although errors in
[Fe/H] better than 0.10~dex are rare, [X/Fe] can often be believed at the level
of 0.05~dex  (though in some cases only 0.15~dex may be achieved). 
Homogeneous studies, where abundances are derived almost identically
for all stars, can achieve better internal accuracy.

Additional errors can 
arise from the assumptions of the
analysis:
\newline $\bullet$ Effective temperature scales are a large
source of error. Infra-red flux
method (IRFM) temperatures, often argued to be preferable to other
photometric calibrations, are now available for many
halo stars (Alonso, Arribas \& Mart\'inez-Roger 1996), although the
uncertainties on any individual star are currently large.
\newline $\bullet$ Collisional damping constants have 
improved to the extent that {\it strong} lines may now
provide more reliable abundances than {\it weak}
lines (Anstee, O'Mara \& Ross 1997). 
Damping constants for many neutral transitions have been
published (Anstee \& O'Mara 1995; Barklem \& O'Mara 1997,1998; 
Barklem, O'Mara \& Ross 1998); comparisons with older computations
are presented elsewhere (Ryan 1998).
\newline $\bullet$ 
Corrections for non-LTE have been
published for several elements: e.g.
Li (Carlsson et al. 1994; Pavlenko \& Magazzu 1996),
Be (Garc\'ia L\'opez, Severino \& Gomez 1995),
B (Kiselman 1994; Kiselman \& Carlsson 1996),
O (Kiselman 1991),
Na (Baum\"uller, Butler \& Gehren 1998),
Mg (Zhao, Butler \& Gehren 1998),
Al (Baum\"uller \& Gehren 1997),
and 
Ba (Mashonkina, Gehren \& Bikmaev 1999).
\newline $\bullet$ 3-D hydrodynamical models are being computed to
investigate the errors introduced by 1-D models. Preliminary 
work signals some interesting results (Asplund et al. 1999), 
including the primordial Li abundance having been overestimated.

\section{The Lightest Elements}

The primordial and spallative elements
Li, Be, and B will be thoroughly considered by IAU Symposium
198.  In the context of
globular clusters versus field stars (see Figure~3), it is useful to
compare M92 subgiants (Boesgaard et al. 1998) and
field halo turnoff stars (Ryan, Norris \& Beers 1999). 
The latter show no intrinsic spread 
($\sigma_{\rm int} < 0.02$~dex) once a small 
metallicity dependence (believed to be due to GCE) is taken
into account, whereas M92 shows a 
range of Li. 
This suggests a difference between
the field and globular cluster populations,  
presumably related to their very different environments, 
specifically the stellar density.
Although single stars seldom interact, protostellar disks may
have done so in
nascent globular clusters but not in lower density clusters destined ultimately 
to produce
field stars (Kraft 1998, private communication). Such speculation falls
short, however, of explaining why the M92 subgiants might have a higher Li
spread than in the field. Boesgaard et al. favoured 
rotationally-induced turbulence resulting in a spread of Li depletion factors
from a higher initial value.
One might
imagine that different angular momentum histories of cluster 
and field stars could lead to differences of this sort, though
the extreme thinness of the field star A(Li)
distribution (Ryan et al. 1999)
is a lingering difficulty with that scenario.

%% \begin{figure}
%% \vspace{5 cm}
%% \caption{Metallicity distributions of Galactic halo globular clusters, 
%% Galactic halo field stars, and Damped Lyman-$\alpha$ quasar absorption line 
%% systems.}
%% \end{figure}

\begin{figure}[!htb]
\begin{center}
\leavevmode
\epsfxsize=150mm
\epsfbox[20 396 564 689]{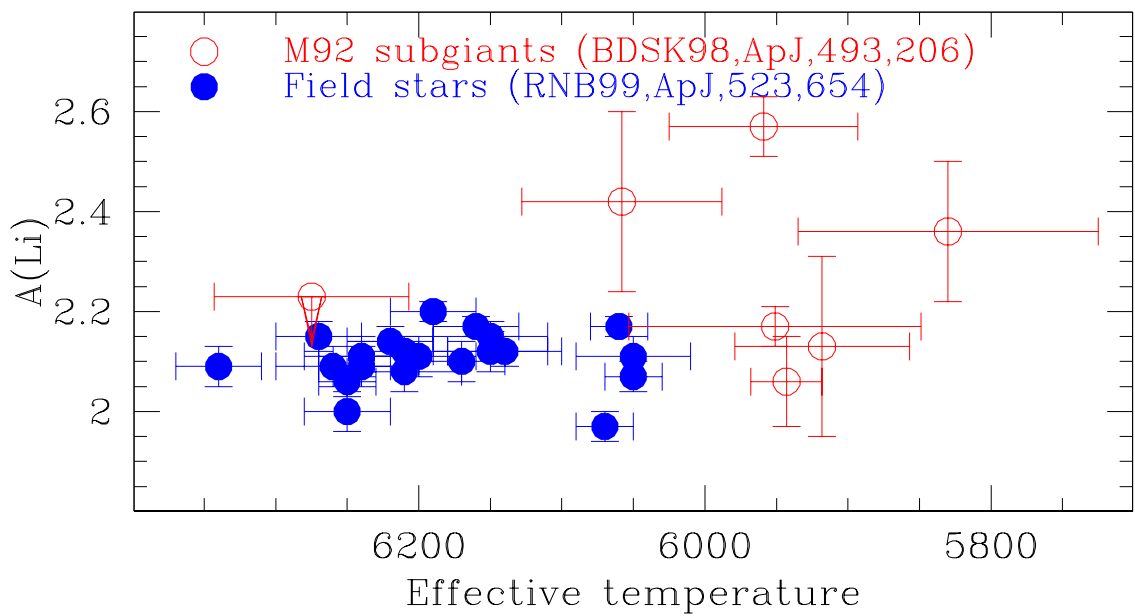}
% \epsfbox{liege_fig3.ps}
\end{center}
% % psfile=#1 vsize=#2 angle=#3 hscale=#4 vscale=#5 hoffset=#6 voffset=#7
% \plotfiddle{sample_fig1.eps}{75truemm}{0}{45}{45}{-150}{-80}
\caption{Lithium abundances in very metal-poor halo field stars near the turnoff,
compared with globular cluster subgiants in M92.}
\end{figure}

\section{Intermediate-Mass Elements}

As two of the most abundant metals, C and O are very important in
stellar evolution and as diagnostics of GCE, but our views
of O have
undergone numerous revisions in the last decade. Measurements are presented in Figure 4, which
isolates giants from dwarfs, and shows separately the abundances derived from 
the forbidden lines [O I], near-infrared
triplet O I, and ultraviolet OH lines.

%% \begin{figure}
%% \vspace{5 cm}
%% \caption{Metallicity distributions of Galactic halo globular clusters, 
%% Galactic halo field stars, and Damped Lyman-$\alpha$ quasar absorption line 
%% systems.}
%% \end{figure}

\begin{figure}[!htb]
\begin{center}
\leavevmode
\epsfxsize=150mm
\epsfbox[148 401 564 707]{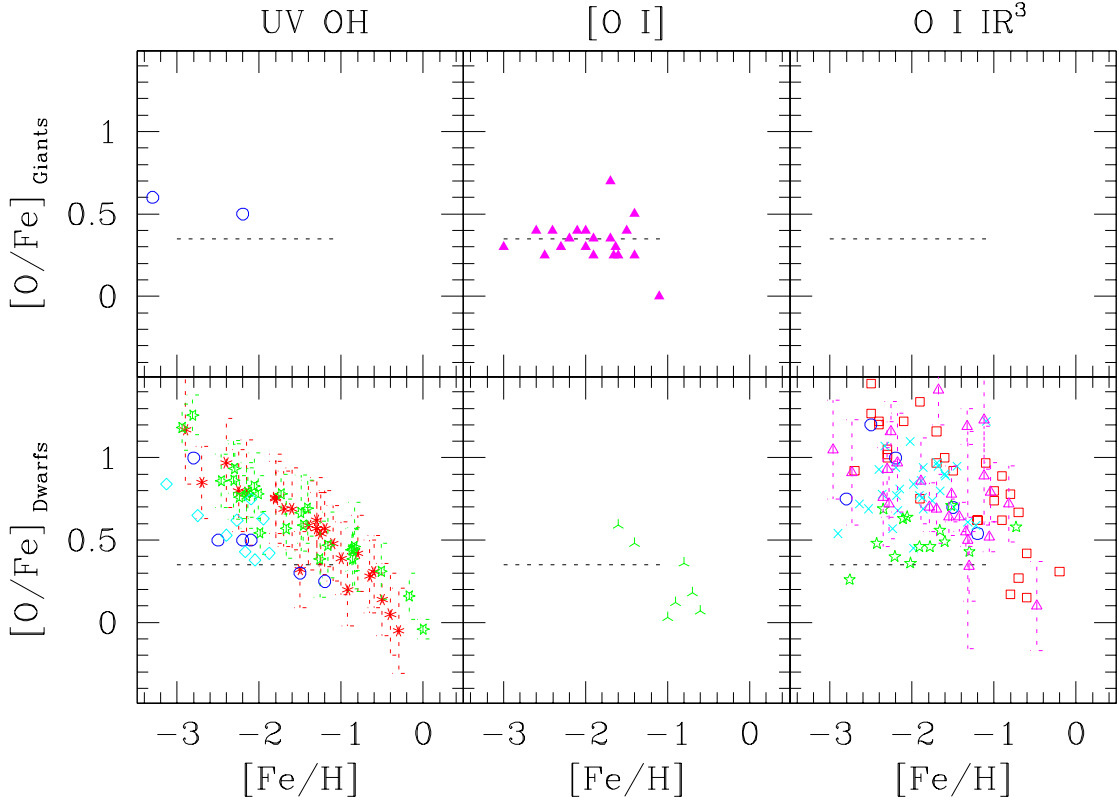}
% \epsfbox{liege_fig3.ps}
\end{center}
% % psfile=#1 vsize=#2 angle=#3 hscale=#4 vscale=#5 hoffset=#6 voffset=#7
% \plotfiddle{sample_fig1.eps}{75truemm}{0}{45}{45}{-150}{-80}
\caption{Oxygen in metal poor field stars. 
{\it circles:} Bessell, Sutherland \& Ruan 1991;
{\it filled triangles:} Barbuy 1988;
{\it diamonds:} Nissen et al. 1994;
{\it six-pointed stars:} Boesgaard et al. 1999;
{\it asterisks:} Israelian, Garc\'ia Lop\'ez \& Rebolo 1998;
{\it inverted `Y's:} Spite \& Spite 1991;
{\it squares:} Abia \& Rebolo 1989;
{\it crosses:} Tomkin et al. 1992;
{\it five-pointed stars:} King 1993;
{\it open triangles:} Cavallo, Pilachowski \& Rebolo 1997.
}
\end{figure}

Barbuy (1988), studying the forbidden 
line in giants,
found a result similar to that for $\alpha$-elements in the halo,
i.e. an overabundance by 0.3--0.4~dex irrespective of metallicity,
shown as a dotted line in Figure 4.
Results for the infra-red triplet, however, have been  
inconsistent. Abia \& Rebolo (1989) found a
progressive increase in the overabundance, approaching 1.5~dex at [Fe/H]~=~$-3$.
Subsequent studies found different results, arguing
inconsistencies between different lines (Magain 1988), $T_{\rm eff}$
dependences, and errors in the $T_{\rm eff}$ scale and equivalent widths. 
The most recent studies used the UV OH lines,
Israelian et al. (1998) and Boesgaard et al. (1999) finding
almost identical trends towards high O overabundances in the most metal-poor
stars.  Note that although Bessell, Sutherland \& Ruan
(1991) concluded that their OH line measurements were consistent with 
Barbuy's (1988) giants, Bessell et al's dwarfs and Nissen et al's (1994)
almost fit the recent OH trend, albeit displaced by $\simeq -0.3$~dex.
Gratton (1999) 
warns of the difficulty of fitting the 
continuum near the UV OH lines for the more
metal-rich stars, the need for accurate molecular line
parameters, and the need to apply NLTE corrections for the 
high-excitation O~I triplet. The lack of agreement between measurements of the
triplet (Figure~4) attests to the difficulties in using these lines.

Barbuy's finding that O behaves like the
$\alpha$-elements was attractive under the paradigm that copious
production of Fe in SN~Ia was responsible for the decrease in [$\alpha$/Fe]
at [Fe/H]~$>$~$-1$. However, O and the $\alpha$ elements are formed separately,
so their yields relative to
Fe do not have to exhibit the same dependence on metallicity. 
The difficulty posed 
by the observations is 
that current stellar models predict
O and $\alpha$ to evolve together,
and do not show high overabundances of [O/Fe].  
The models of Timmes, Woosley \& Weaver (1995), for example, achieve
[O/Fe]~=~+0.4 at [Fe/H]~$\simeq$~$-3$, or even +0.6 if the iron yield is
halved, but they don't reach 1.0 as the OH data do.
Furthermore, it is not sufficient to claim that the Fe yields
could be wrong, for while they {\it could} be wrong, that would not help
the problem with O; the OH observations require that [O/$\alpha$] is also
strongly dependent upon metallicity, and that is not seen in the models
either. So, if the theoretical yields are to fit the OH line data, significantly
higher O production will be required, irrespective of what changes
are made to Fe.

Carbon and nitrogen measurements in very metal-poor stars
(Israelian \& Rebolo 1999) generally give
approximately solar ratios.  
However,
there 
have also been
significant numbers of C-rich stars found at the lowest iron
abundances (Beers, Preston \& Shectman 1992). Possibly as many as 10\% of stars
with [Fe/H]~$<$~$-2$ have high CH-band strengths.
While one of these,
CS~22892-052,  has huge r-process
element overabundances (Sneden et al. 1994, 1996), the majority 
exhibit s-process patterns (Norris, Ryan \& Beers 1997a; Barbuy et al. 1997;
Norris et al. 2000).
The high C abundances are not predicted by theoretical yields of supernovae 
(e.g. Timmes et al. 1995), but might be explained by enrichment of
the early halo by mass loss from high mass 
Population III stars  
which have synthesized C via the triple-alpha process and mixed it to their 
surfaces (e.g. Marigo 1999). 
The frequency of C-rich stars and 
their tendency to be accompanied by s-process rather than r-process heavy
element signatures (suggestive of AGB star evolution rather than 
supernovae)  will be important to understanding their origin.
(Note, however, that not all C-rich stars exhibit heavy element anomalies
[Norris, Ryan \& Beers 1997b]).

The $\alpha$-elements (Mg, Si, Ca) have fairly uniform
overabundances relative to iron extending down to the lowest metallicities
(Ryan et al. 1991; Norris et al. 1993; McWilliam et al. 1995; 
Ryan et al. 1996). 
Recent studies have begun
to concentrate on star-to-star variations, with King (1997),
Carney et al. (1997), and Nissen \& Schuster (1997) identifying
stars with low [$\alpha$/Fe] ratios,
predominantly retrograde kinematics, and young ages. The low
[$\alpha$/Fe] ratios are reminiscent of those proposed by 
Matteucci \& Brocatto (1990) and 
Gilmore \& Wyse (1991) for star formation in dwarf galaxies where star formation
ceased before metallicities typical of Galactic disk stars were reached, so that
the appearance of SN~Ia would lead to low [$\alpha$/Fe] ratios even at 
metallicities typical of halo stars.\footnote{A possible conflict 
is raised by the work of Kobayashi et al. (1998) who propose that the
SN~Ia mechanism would have lower efficiency at
[Fe/H]~$<$~$-1$
due to the weaker winds in metal-deficient stars.}
Similarly, the dual halo models of
Zinn (1993), based on globular clusters, 
and Norris (1994), based on field stars,
fit with the characteristics of the stars now observed. Clearly there is still
much to learn about the Galactic halo by exploiting 
relative abundances.

For [Na/Fe] and [Mg/Fe], Hanson et al. (1998) have found 
different behaviours 
for globular cluster and field stars, even though they
have examined the same evolutionary states in each sample. 
They find the two elements to be positively correlated in the field, signifying
common production, and find correlations with kinematics in the sense that
the youngest stars (inferred from lower [Na/Fe] and 
[Mg/Fe] values) have predominantly retrograde kinematics. In clusters, on the
other hand, they find overall an anti-correlation of [Na/Fe] and [Mg/Fe]
suggestive of the proton-capture chain having been active (converting
Ne to Na, and Mg to Al, during H-burning via the CNO cycle). The appearance at
the stellar surface of these changes requires 
deep mixing, so one may conclude that deep mixing has
occurred in the globular cluster giants but not in the field giants 
{\it even at the same evolutionary state}. Why this should be so has yet to be
resolved; the Li results (\S 4) may be related.

\section{The Iron-Peak Elements}

The iron-peak elements most easily observed in very metal-poor stars are
Sc, Ti, Cr, Mn, Fe, Co, and Ni. For stars with [Fe/H]~$>$~$-2.5$, most of these
species exhibit solar abundance ratios. One exception is
Mn, which is underabundant by $\sim$0.3~dex in halo stars. A second exception
is Ti, but whether or not Ti 
should be included in the iron-peak group at all is unclear 
(e.g. Lambert 1987). While it is believed to be produced 
in the same region of stars as the other iron-peak elements, its abundance 
appears more like the $\alpha$-elements in that it is overabundant 
by $\sim$0.3--0.4~dex in stars with [Fe/H]~$<$~$-1$.\footnote{Truran (1999, 
private communication) reports that the experimental 
$^{44}$Ti($\alpha$,p)$^{47}$V rate is five times higher than the theoretical
one used in previous nucleosynthesis calculations, so the theoretical yields
of some iron-peak elements will be subject to revision.}
McWilliam et al. (1995) and Ryan et al. (1996) showed that although the Sc and
Ti 
abundance trends
persist to the lowest metallicities known, [Fe/H]~=~ $-4.0$, 
Mn and Cr become very underabundant in stars with [Fe/H~$<$~$-2.5$, while in
the same objects Co becomes overabundant. (See Figure 5.)
There is also limited
evidence for Ni overabundances in some of these objects. 
These changes were not predicted by supernova computations, and provide
a recent example of observations leading theory in new directions.

%% \begin{figure}
%% \vspace{5 cm}
%% \caption{Metallicity distributions of Galactic halo globular clusters, 
%% Galactic halo field stars, and Damped Lyman-$\alpha$ quasar absorption line 
%% systems.}
%% \end{figure}

\begin{figure}[!htb]
\begin{center}
\leavevmode
\epsfxsize=150mm
\epsfbox[33 167 525 690]{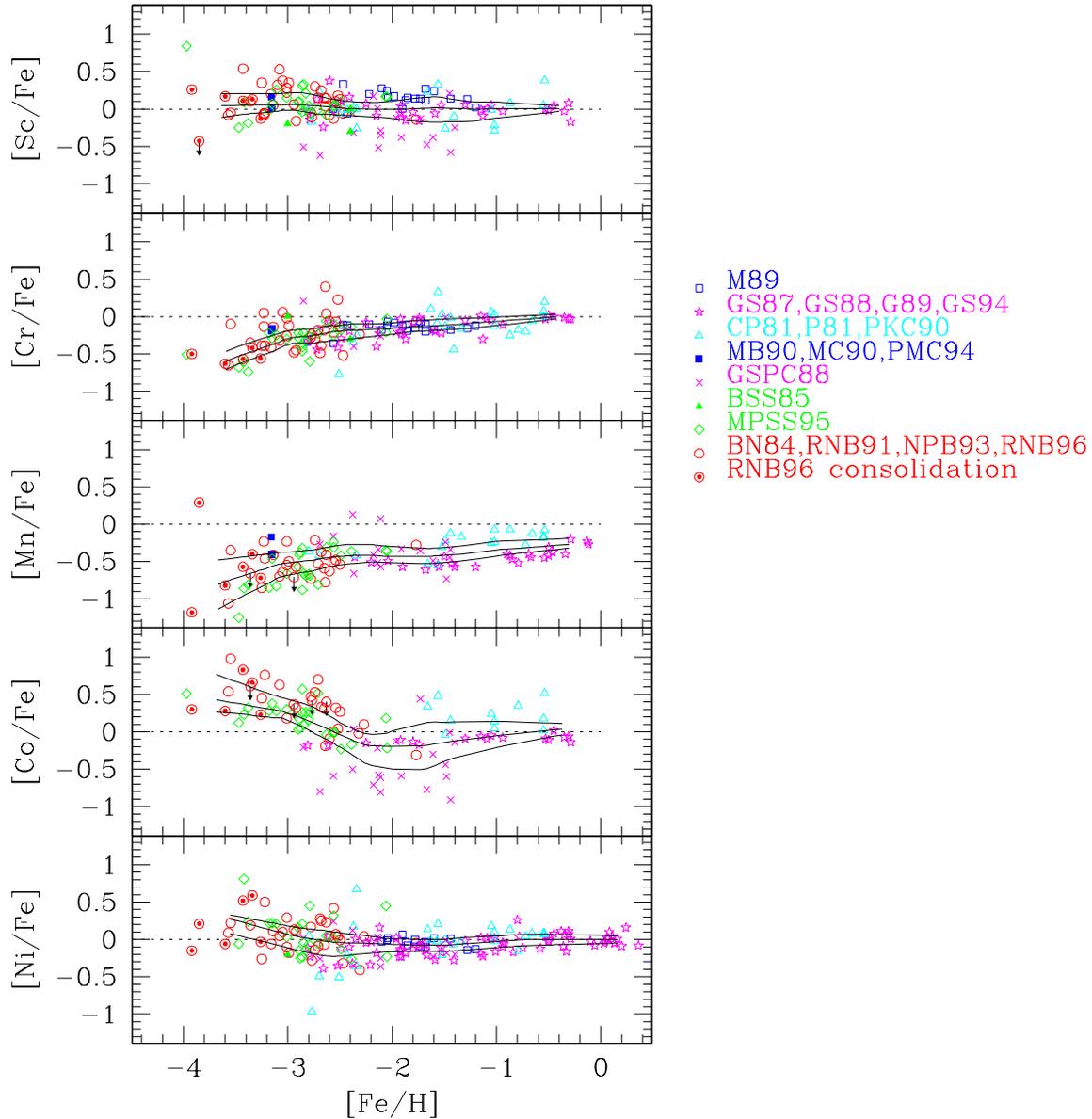}
% \epsfbox{sccrmnconife.ps}
\end{center}
% % psfile=#1 vsize=#2 angle=#3 hscale=#4 vscale=#5 hoffset=#6 voffset=#7
% \plotfiddle{sample_fig1.eps}{75truemm}{0}{45}{45}{-150}{-80}
\caption{Abundances of iron-peak elements in very metal-poor halo field stars.
Solid curves are robust estimates of the mean and quartiles.
(See text and references for details.)}
\end{figure}

The iron-peak species are synthesised deep in the 
star, close to mass-cut (the division between matter
ejected from the supernova and that which collapses onto the stellar remnant).
Supernova calculations are unable to eject material
naturally from the physical conditions
of the star, foreshadowing the difficulty of accurately
predicting the yields of the iron-peak species which depend
sensitively on the explosive conditions. The
position of the mass cut cannot be predicted, but must be constrained
by the observed abundances.
Cr and Mn are produced only in a shell towards the outside of the Fe (and Ti)
core, while Co and Ni are produced internal to that, so the resulting
relative yields of iron-peak species can be adjusted by moving the 
mass-cut even slightly within the star (e.g. Nakamura et al. 
1999). 

Efforts to understand the abundance trends have 
focussed on the possible $\alpha$-rich freeze-out
(Woosley \& Hoffman 1992; McWilliam et al. 1995), the location of the mass-cut
(Nakamura et al. 1999), and the dependence of yields on
stellar mass, metallicity, and neutron excess (Nakamura et al. 1999; 
Hix \& Thielemann 1996). The explanation must account for
a handful of elements with different behaviours, whilst not
distorting the [X/Fe] ratios of species such as Mg, Si, and Ca which are
formed well outside the mass-cut. This is particularly challenging 
for explanations which alter the Fe yield!
Progress may increase when 3D 
hydrodynamical models provide more realistic treatments of the 
supernova explosion 
(e.g. M\"uller, Fryxell \& Arnett 1991). 
Observational evidence for mixing of material from the mass-cut to the 
photosphere appears to be growing 
(Stathakis 1996; Fassia et al. 1998), and provides an additional
set of clues and 
constraints as we seek more realistic supernova models.

\section{Neutron-Capture Elements}

Spite \& Spite (1978)
showed from observations of Ba (of mixed s- and r-process origin) and
Eu (essentially pure r-process) that the Ba in halo stars owed its
origin primarily to the r-process. Truran (1981) provided a theoretical insight into 
the relative contributions of the processes by examining the roles of 
seed
nuclei, and argued that the s-process should be viewed as a secondary process
while the r-process was primary, and thus that 
first generation --- Population III --- stars would not
execute the s-process. 
Gilroy et al. (1988) showed that the neutron-capture elements exhibited 
abundance patterns more consistent with the r-process than
the s-process. 
McWilliam et al. (1995, 1998)
extended the Ba and Eu comparison to the lowest metallicities  
([Fe/H]~=~$-4$), confirming the existence of r-process (rather than s-process)
ratios. 

An alternative means of examining the s- and r-process contributions
would be via isotope ratios. 
Heavy element isotope lines are invariably blended in stellar
spectra, but in a small number of cases they give rise to 
significant differences in the line profile allowing constraints
to be placed on the isotope ratios. Magain (1995) obtained observations of this
effect, and in HD~140283, at [Fe/H]~=~$-2.6$, inferred an s-process
rather than r-process pattern for Ba. This result was unexpected given the
r-process framework that had been established over the preceeding nearly 20
years. In contrast, Gacquer \& Francois (1998, private communication)  find
an r-process signature for this star.

The reliance almost solely on Ba and Eu as 
diagnostics of the s- vs r-process fractions 
has been diminished by {\it HST} data for other
important neutron-capture elements having UV spectra. Included in this list
are Ag, Pt, Os, and Pb (Crawford et al. 1998; Cowan et al. 1996; 
Sneden et al. 1998).

Gilroy et al. (1988) also found that neutron-capture elements showed
significant star-to-star variations in the most metal poor objects
([Fe/H]~$^<_\sim$~$-2.5$), building upon similar cases 
reported earlier during the decade. 
The abundance variations are
greatest for Sr, where ranges of a factor of more than 100 were
found amongst dwarfs (Ryan et al. 1991) and giants (Norris et al. 1993).
Moreover, as Figure~6 shows, such extreme variations are {\it not}
shared by Ba, which
exhibits a fairly steady trend towards lower [Ba/Fe] at lower [Fe/H].

%% \begin{figure}
%% \vspace{5 cm}
%% \caption{Metallicity distributions of Galactic halo globular clusters, 
%% Galactic halo field stars, and Damped Lyman-$\alpha$ quasar absorption line 
%% systems.}
%% \end{figure}

\begin{figure}[!htb]
\begin{center}
\leavevmode
\epsfxsize=150mm
\epsfbox[32 407 526 689]{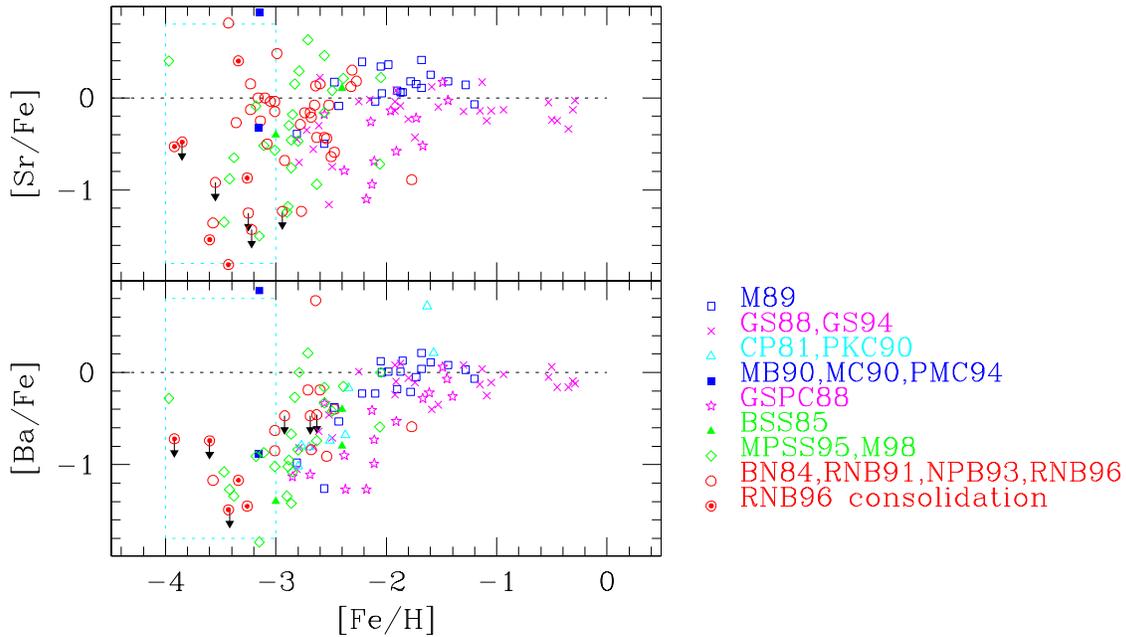}
% \epsfbox{liege_fig5.ps}
\end{center}
% % psfile=#1 vsize=#2 angle=#3 hscale=#4 vscale=#5 hoffset=#6 voffset=#7
% \plotfiddle{sample_fig1.eps}{75truemm}{0}{45}{45}{-150}{-80}
\caption{Abundances of Sr and Ba in halo stars. The dashed box outlines the
same region of [X/Fe] and [Fe/H] in each plot. Whereas [Sr/Fe] exhibits a range
by a factor of more than 100, [Ba/Fe] has a well-behaved trend towards lower
values at the lowest [Fe/H]. }
\end{figure}

Reconciling Sr and Ba is a challenge. If the 
r-process alone is responsible for their synthesis 
in very metal-poor halo stars, then Figure~6 tells us that the r-process
cannot be universal. Whilst there is no theoretical reason why it {\it must}
be universal, halo star neutron-capture element abundance
patterns generally resemble the 
r-process contribution in the sun (e.g. Gilroy et al. 1988; 
McWilliam et al. 1995; Cowan et al. 1995, Sneden et al. 1996). This cannot
be ignored, even if non-uniqueness (Goriely \& Arnould 1997) is 
possible. Hypothesising that the lower envelope to the [Sr/Fe] 
observations is the ``normal'' r-process behaviour,  
corresponding to a universal [Ba/Sr] r-process value, we seek a source
of additional Sr in the lowest metallicity objects.
Figure~6 emphasises that the process responsible
must involve low
neutron exposures that synthesise only species around the atomic number of Sr.
Other species near Sr are also enhanced in these
high Sr stars (Ryan et al. 1996, 2000; see Figure~7.) 
Whilst the weak s-process would 
produce primarily low atomic-numbered neutron-capture species, and would be
active in normal high mass ($>$15~M$_\odot$) stars, the difficulties 
related to the lack of
seed nuclei and a suitable neutron source would severely hamper this
in low metallicity stars. It may be necessary 
to
look for a new, low-neutron-exposure r-process (Ishimaru \& Wanajo
1999b), reflecting two different types of core collapse
supernovae.
% depending on whether the stellar remnant is a black hole or neutron star. 

\begin{figure}[!htb]
\begin{center}
\leavevmode
\epsfxsize=150mm
\epsfbox[32 439 564 689]{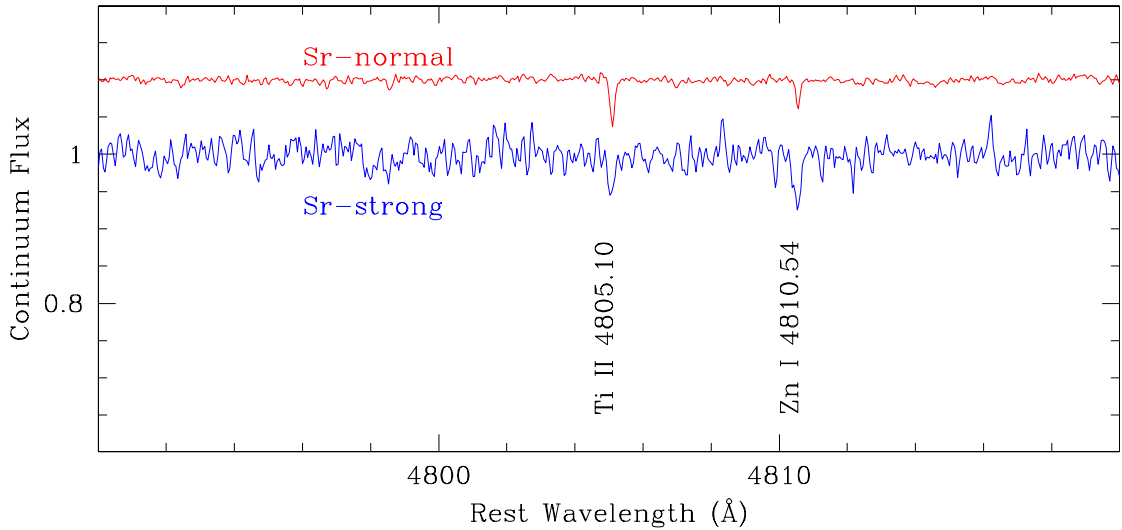}
% \epsfbox{liege_fig5.ps}
\end{center}
% % psfile=#1 vsize=#2 angle=#3 hscale=#4 vscale=#5 hoffset=#6 voffset=#7
% \plotfiddle{sample_fig1.eps}{75truemm}{0}{45}{45}{-150}{-80}
\caption{Spectra of Fe and Zn lines in Sr-rich (lower) and Sr-normal (upper) 
stars. The 
investigation of other low-atomic-number neutron-capture elements indicates
that these too are overabundant in the Sr-rich stars, as expected for a
low-neutron-exposure process (Ryan et al. 2000).}
\end{figure}

GCE models need to reproduce the climb in 
[Ba/Fe] and [Eu/Fe] for stars with [Fe/H]~$<$~$-2$, which suggests
that these species are not produced abundantly in the most massive
metal-poor supernovae, for if they were then they would exhibit higher ratios
(to iron) even in the lowest metallicity stars 
(e.g. Mathews \& Cowan 1990). 
Mathews \& Cowan
associated their production
with 10--11~M$_\odot$ stars.  Although the evolution 
in their models is now seen to be too steep for today's data, the
concept of low-mass stars being responsible has survived, and a more
gradual emergence is seen in the models of Travaglio et al. (1999) which
use a mass range 8--10~M$_\odot$ for the process. Travaglio
et al. also treat the 
enrichment of the cloud inhomogeneously, along the lines described 
in
\S2 of this review (also Ishimaru \& Wanajo 1999a).

The final comments on the neutron-capture elements are reserved for the
atypical r-process-rich star, CS~22892-052. Assuming that the neutron-capture
elements in this star are present in the same proportions as the r-process 
contribution in the sun,
Cowan et al. (1997) derive an age of 17$\pm$4~Gyr, based on
the radioactive Th abundance. This is consistent with current estimates
of globular cluster ages from isochrone fitting, of around 12.8$^{+4.2}_{-2.8}$~Gyr 
(Chaboyer 1999).
Improved data and GCE models will hopefully reduce the errors
to provide a chronometer of better accuracy than the
globular cluster technique.

\section{Final Comments and Summary}

Stars with [Fe/H]~$<$~$-3$ formed at the earliest epochs of star formation, 
corresponding to redshifts $z~^>_\sim~4$.
They formed from gas clouds enriched by single supernovae, and hence allow
us to
investigate the first (Population III) stars which enriched primordial
(big bang) material with heavy elements. Even though that stellar
population was small in number and no surviving example has been detected
today, it is still possible to investigate its mass function and evolution
via the nucleosynthetic yields frozen into the next generation of stars,
the extremely metal-poor Population II stars.

Extremely metal-poor stars exist in two main environments --- the field and
globular clusters --- and the two appear not to be identical. Building on
previous differences (e.g. CN variations which appear more common in
globular cluster than field stars), we now see a spread
of Li in M92 and evidence of deep mixing
in globular cluster giants (via the proton-chain signature)
which do not appear in the field. These may indicate that the
clusters in which today's field stars formed were not the same as the dense
globular clusters that survive to the present. In using Population II objects
to study GCE, it would then seem safer to rely on field stars which sample a
much larger volume of the Galaxy and are less likely to have interacted with
other objects in a dense environment, rather than trusting objects
which sample the chemical evolution of 
a small region of space with high stellar density. 

Recent measurements have highlighted the state of confusion which
exists over oxygen abundances in halo stars. 
Meanwhile,
some elements have shown remarkably robust abundance patterns as 
more metal-deficient stars are sampled, such as Ti (though we know not why!).
Greater accuracy has enabled star-to-star abundance variations to be examined
in intermediate atomic-mass elements, shedding new light on the chemical and
kinematic composition of the Galaxy, and its evolution. Likewise, the huge
abundance spread in Sr, but not Ba, is allowing us to probe
the nucleosynthesis processes and ultimately the mass function of the
first stars.
These advances accompany more realistic GCE models that incorporate the
stochastic nature of star formation which is potentially of great importance in the
earliest epochs where small numbers of supernovae were involved.

\beginrefer

\refer Abia, C. \& Rebolo, R. 1989, ApJ,  { 347}, 186

\refer Alonso, A., Arribas, S. \& Mart\'inez--Roger, C. 1996, A\&AS, 117, 227

\refer Anstee, S. D. \& O'Mara, B. J. 1995, MNRAS, 276, 859

\refer Anstee, S. D., O'Mara, B. J. \& Ross, J. E. 1997, MNRAS, 284, 202

\refer Argast, D. \& Samland, M. 1999, The Galactic Halo: from Globular Clusters
to Field Stars, 35$^{\mbox{\rm th}}$ Li\`ege\ Int.\ Astroph.\ Coll., in press

\refer Asplund, M., Nordlund, \AA., Trampedach, R. \& Stein, R. F. 1999, astro-ph/9905059

\refer Audouze, J. \& Silk, J. 1995, ApJ, 451, L49

\refer Barbuy, B. 1988, A\&A,  191, 121

\refer Barbuy, B.  Cayrel, R., Spite, M., Beers, T. C., Spite, F., Nordstrom, B. \& Nissen, P. E. 1997, A\&A, 317, L63

\refer Barbuy, B., Spite, F., \& Spite, M. 1985, A\&A, 144, 343

\refer Barklem, P. S. \& O'Mara, B. J. 1997, MNRAS, 290, 102

\refer Barklem, P. S. \& O'Mara, B. J. 1998, MNRAS, 300, 863

\refer Barklem, P. S., O'Mara, B. J. \& Ross, J. E. 1998, MNRAS, 296, 1057

\refer Baum\"uller, D., Butler, K. \& Gehren, T. 1998, A\&A, 338, 637

\refer Baum\"uller, D. \& Gehren, T. 1997, A\&A, 325, 1088

\refer Beers, T. C., Preston, G.  W. \& Shectman, S.  A. 1992,  AJ, 103, 1987

\refer Bessell, M. S. \& Norris, J. E. 1984, ApJ,  { 285}, 622

\refer Bessell, M. S., Sutherland, R. S. \& Ruan, K. 1991, ApJ, 383, L71

\refer Boesgaard, A. M., Deliyannis, C. P., Stephens, A. \& King, J. R. 1998, ApJ, 493, 206

\refer Boesgaard, A. M., King, J. R., Deliyannis, C. P. \& Vogt, S. S. 1999, AJ, 117, 492

\refer Bouwens, R., Broadhurst, T., \& Silk, J. 1998, ApJ, 506, 557

\refer Carlsson, M., Rutten, R. J., Bruls, J. H. M. J. \& Shchukina, N. G. 1994, A\&A, 288, 860

\refer Carney, B. W. \& Peterson, R. C. 1981, ApJ,  { 245}, 238

\refer Carney, B. W., Wright, J. S., Sneden, C., Laird, J. B., Aguilar, L. A. \& Latham, D. W. 1997, AJ, 114, 363

\refer Cavallo, R. M., Pilachowski, C. A. \& Rebolo, R. 1997, PASP, 109, 226

\refer Chaboyer, B. 1999,  The Galactic Halo: from Globular Clusters
to Field Stars, 35$^{\mbox{\rm th}}$ Li\`ege\ Int.\ Astroph.\ Coll., in press

\refer Cioffi, D. F., McKee, C. F. \& Bertschinger, E. 1988, ApJ, 334, 252

\refer Cowan, J. J., Burris, D. L., Sneden, C., McWilliam, A. \& Preston, G. W. 1995, ApJ, 439, L51

\refer Cowan, J. J., McWilliam, A., Sneden, C., \& Burris, D. L. 1997, ApJ, 480, 246

\refer Cowan, J. J., Sneden, C., Truran, J. W. \& Burris, D. L. 1996, ApJ, 460, L115

\refer Crawford, J. L., Sneden, C., King, J. R., Boesgaard, A. M. \& Deliyannis, C. P.  1998, AJ, 116, 2489

\refer Edmunds, M. G. \& Phillipps, S. 1997, MNRAS, 292, 733

\refer Edvardsson, B., Andersen, J., Gustafsson, B., Lambert, D. L., Nissen, P. E. \& Tomkin, J. 1993, A\&A, 275, 101

\refer Fassia, A., Meikle, W. P. S., Geballe, T. R., Walton, N. A., Pollaco, D. L., Rutten, R. G. M. \& Tinney, C. 1998, MNRAS, 299, 150

\refer Garc\'ia L\'opez, R. J., Severino, G. \& Gomez, M. T. 1995, A\&A, 297, 787

\refer Gilmore, G. \& Wyse, R. F. G. 1991, ApJ, 367, L55

\refer Gilroy, K. K., Sneden, C., Pilachowski, C. A. \& Cowan, J. J. 1988, ApJ, 327, 298

\refer Goriely, S. \& Arnould, M. 1997, A\&A, 322, L29

\refer Gratton, R. G. 1989, A\&A, { 208}, 171

\refer Gratton, R. 1999, The Galactic Halo: from Globular Clusters
to Field Stars, 35$^{\mbox{\rm th}}$ Li\`ege\ Int.\ Astroph.\ Coll., in press

\refer Gratton, R. G. \&  Sneden, C. 1987, A\&A, { 178}, 179

\refer Gratton, R. G. \&  Sneden, C. 1988, A\&A, 204, 193

\refer Gratton, R. \& Sneden, C. 1994, A\&A, 287, 927

\refer Hanson, R. B., Sneden, C., Kraft, R. P. \& Fulbright, J. 1998, AJ, 116, 1286

\refer Hellsten, U., Dave, R., Hernquist, L., Weinberg, D. H. \& Katz, N. 1997, ApJ, 487, 482

\refer Hix, W. R. \& Thielemann, K.-F. , 1996, ApJ, 460, 869

\refer Ishimaru, Y. \& Wanajo, S. 1999a, ApJ, 511, L33

\refer Ishimaru, Y. \& Wanajo, S. 1999b, MPA/ESO Workshop on the First Stars,
Garching 1999, in press.

\refer Israelian, G., Garc\'ia L\'opez, R. \& Rebolo, R. 1998, ApJ, 507, 805

\refer Israelian, G. \& Rebolo, R. 1999, The First Stars, MPA/ESO Workshop, Garching August 1999, in press

\refer King, J. R. 1993, AJ, 106, 1206

\refer King, J. R. 1997, AJ, 113, 2302

\refer Kiselman, D. 1991, A\&A, 245, L9

\refer Kiselman, D. 1994, A\&A, 286, 169

\refer Kiselman, D. \& Carlsson, M. 1996, A\&A, 311, 680

\refer Kobayashi, C., Tsuijmoto, T., Nomoto, K. Hachisu, I. \& Kato, M. 1998, ApJ, 503, L155

\refer Laird, J. B., Carney, B. W. \& Latham, D. W. 1988b,  AJ,  { 95}, 1843

\refer Laird, J. B., Rupen, M. P., Carney, B. W.  \& Latham, D. W. 1988a, AJ,  96, 1908

\refer Lambert, D. L. 1987, JApA, { 8}, 103

\refer Lu, L., Sargent, W. L. W., Barlow, T. A., Churchill, C. W. \& Vogt, S. S. 1996, ApJS, 107, 475

\refer Magain, P. 1988, The Impact of Very High S/N Spectroscopy on Stellar 
Physics, ed. G. Cayrel de Strobel \& M. Spite (Dordrecht, Kluwer), 485

\refer Magain, P. 1989, A\&A, 209, 211

\refer Magain, P. 1995, A\&A, 297, 686

\refer Marigo, P. 1999, MPA/ESO Workshop on The First Stars, Garching August 1999, in press

\refer Mashonkina, L., Gehren, T. \& Bikmaev, I. 1999, A\&A, preprint

\refer Mathews, G. J. \& Cowan, J. J. 1990, Nature, 345, 491

\refer Matteucci, F. \& Brocato, E. 1990, ApJ, 365, 539

\refer McWilliam, A., 1998, AJ, 115, 1640

\refer McWilliam, A., Preston, G. W., Sneden, C. \& Searle, L. 1995, AJ, 109, 2757

\refer Molaro, P. \& Bonifacio, P. 1990, A\&A, 236, L5

\refer Molaro, P. \& Castelli, F. 1990, A\&A, { 228}, 426

\refer Molaro, P., D'Odorico, S., Fontana, A., Savaglio, S. \& Vladilo, G. 1996, A\&A, 308, 1

\refer M\"uller, E., Fryxell, B. \& Arnett, D. 1991, A\&A, 251, 505

\refer Nakamura, T., Umeda, Nomoto, K., Thielemann, F.-K. \& Burrows, I. 1999, ApJ, in press (astro-ph/9809307)

\refer Nissen, P. E., Gustafsson, B., Edvardsson, B., \& Gilmore, G. 1994, A\&A, 285, 440

\refer Nissen, P. E. \& Schuster, W. M. 1997, A\&A, 326, 751

\refer Norris, J. E. 1994, ApJ, 431, 645

\refer Norris, J. E., Peterson, R. C. \& Beers, T. C. 1993, ApJ, 415, 797

\refer Norris, J. E., Ryan, S. G. \& Beers, T. C. 1997a, ApJ, 488, 350

\refer Norris, J. E., Ryan, S. G. \& Beers, T. C. 1997b, ApJ, 489, L169

\refer Norris, J. E., Ryan, S. G., Beers, T. C. \& Aoki, W. 2000, in prep.

\refer Nulsen, P. E. J., Barcons, X. \& Fabian, A. C. 1998, MNRAS, 301, 168

\refer Pagel, B. E. J. \& Patchett, B. E. 1975, MNRAS, { 172}, 13

\refer Pavlenko, Ya. V. \& Magazzu, A. 1996, A\&A, 311, 961

\refer Peterson, R. 1981, ApJ,  { 244}, 989

\refer Peterson, R., Kurucz, R. L., \& Carney, B. W. 1990, ApJ,  { 350}, 173

\refer Pettini, M., Boksenberg, A. \& Hunstead, R. W. 1990, ApJ, 348, 48

\refer Pettini, M., Ellison, S. L., Steidel, C. C. \& Bowen, D. V. 1999a, ApJ, 510, 576

\refer Pettini, M., Smith, L. J., King, D. L. \& Hunstead, R. W. 1997, ApJ, 486, 665

\refer Pettini, M., Steidel, C. C., Adelberger, K. L., Dickinson, M. \& Giavalisco, M. 1999b, preprint

\refer Primas, F., Molaro, P., \& Castelli, F. 1994, A\&A, 290, 885

\refer Ryan, S. G. 1996, ASP Conf.Ser. 92, 113

\refer Ryan, S. G. 1998, A\&A, 331, 1051

\refer Ryan, S. G., Blake, L., Norris, J. E. \& Beers, T. C. 2000, in preparation

\refer Ryan, S. G. \& Norris, J. E. 1991, AJ, { 101}, 1865

\refer Ryan, S. G., Norris, J. E. \& Bessell, M. S. 1991, AJ, { 102}, 303

\refer Ryan, S. G., Norris, J. E., \& Beers, T. C. 1996, ApJ, 471, 254

\refer Ryan, S. G., Norris, J. E. \& Beers, T. C. 1999, ApJ, 523, 654

\refer Sadler, E. M., Rich, R. M. \& Terndrup, D. M. 1996, AJ 112, 171

\refer Searle, L.  1977 In: Tinsley, B. M. \& Larson, R. B. (eds)
The Evolution of Galaxies and Stellar Populations,
Yale Uni. Obs., New Haven, p.219

\refer Searle, L. \& Sargent, W. L. W. 1972, ApJ, { 173}, 25

\refer Shigeyama, T. \& Tsujimoto, T. 1998, ApJ, 507, L135

\refer Sneden, C., Cowan, J. J., Burris, D. L. \& Truran, J. W. 1998, ApJ, 496, 235

\refer Sneden, C., Preston, G. W., McWilliam, A. \& Searle, L. 1994, ApJ, 431, L27

\refer Sneden, C., McWilliam, A., Preston, G. P., Cowan, J. J., Burris, D. L. \& Armosky, B. J., 1996, ApJ, 467, 819

\refer Spite, M. \& Spite, F. 1978, A\&A, 67, 23

\refer Spite, M. \& Spite, F. 1991, A\&A, 252, 689

\refer Stathakis, R. A. 1996, PhD Thesis, Sydney Uni.

\refer Steidel, C. C., Adelberger, K. L., Giavalisco, M., Dickinson, M. \& Pettini, M., 1999, astro-ph/9811399

\refer Timmes, F. X., Woosley, S. E., \& Weaver, T. A. 1995, ApJS, 98, 617

\refer Tomkin, J., Lemke, M., Lambert, D. L. \&\ Sneden, C. 1992, AJ, 104, 1568

\refer Travaglio, C., Galli, D., Gallino, R., Busso, M., Ferrini, F. \& Straniero, O. 1999, ApJ, 521, 691

\refer Truran, J. W. 1981, A\&A, 97, 391

\refer Tsujimoto, T. \& Shigeyama, T. 1998, ApJ, 508, L151

\refer Warren, S. J., Hewett, P. C. \& Osmer, P. S. 1994, ApJ, 421, 412

\refer Wolfe, A. M., Turnshek, D. A., Smith, H. E. \& Cohen, R. D. 1986, ApJS, 61, 249

\refer Woosley, S. E. \& Hoffman, R. D. 1992, ApJ, 395, 202

\refer Zhao, G., Butler, K. \& Gehren, T. 1998, A\&A, 333, 219

\refer Zinn, R. 1993, ASP.Conf.Ser, 48, 38

\endrefer           
\end{document}